\documentclass[10pt,twocolumn,letterpaper]{article}

\usepackage{iccv}
\usepackage{times}
\usepackage{epsfig}
\usepackage{graphicx}
\usepackage{amsmath}
\usepackage{amssymb}
\usepackage{tabularx}
\usepackage{multirow}
\usepackage{booktabs}
\usepackage{slashed}
\usepackage{subfigure}
\usepackage[table]{xcolor}
\definecolor{dino}{RGB}{249,231,227}

\usepackage[pagebackref=true,breaklinks=true,letterpaper=true,colorlinks,bookmarks=false]{hyperref}

\iccvfinalcopy 

\def\iccvPaperID{****} 

\ificcvfinal\fi

\begin{document}

\title{Cross-Modal Translation and Alignment for Survival Analysis}

\author{Fengtao Zhou$^{1}$ \qquad Hao Chen$^{1, 2, *}$\\
$^{1}$Department of Computer Science and Engineering \\
$^{2}$Department of Chemical and Biological Engineering\\
The Hong Kong University of Science and Technology\\
{\tt\small fzhouaf@connect.ust.hk\qquad jhc@cse.ust.hk}
}

\maketitle
\ificcvfinal\thispagestyle{empty}\fi

\begin{abstract}
   With the rapid advances in high-throughput sequencing technologies, the focus of survival analysis has shifted from examining clinical indicators to incorporating genomic profiles with pathological images. However, existing methods either directly adopt a straightforward fusion of pathological features and genomic profiles for survival prediction, or take genomic profiles as guidance to integrate the features of pathological images. The former would overlook intrinsic cross-modal correlations. The latter would discard pathological information irrelevant to gene expression. To address these issues, we present a Cross-Modal Translation and Alignment (CMTA) framework to explore the intrinsic cross-modal correlations and transfer potential complementary information. Specifically, we construct two parallel encoder-decoder structures for multi-modal data to integrate intra-modal information and generate cross-modal representation. Taking the generated cross-modal representation to enhance and recalibrate intra-modal representation can significantly improve its discrimination for comprehensive survival analysis. To explore the intrinsic cross-modal correlations, we further design a cross-modal attention module as the information bridge between different modalities to perform cross-modal interactions and transfer complementary information. Our extensive experiments on five public TCGA datasets demonstrate that our proposed framework outperforms the state-of-the-art methods. The source code has been released \footnotemark[2].
\end{abstract}

\footnotetext[1]{Corresponding author}
\footnotetext[2]{\href{https://github.com/FT-ZHOU-ZZZ/CMTA}{https://github.com/FT-ZHOU-ZZZ/CMTA}}
\section{Introduction}
Survival analysis is a crucial topic in clinical prognosis research, which aims to predict the time elapsed from a known origin to an event of interest, such as death, relapse of disease, and development of an adverse reaction. Accurate survival prediction is essential for doctors to assess the clinical outcomes for disease progression and treatment efficiency. Traditionally, survival analysis relies on short-term clinical indicators~\cite{hagar2014survival, yu2021dynamic} and long-term follow-up reports~\cite{ahearne2016short, capra2017assessing}, which are time-consuming and impractical in clinical applications. In recent years, medical image analysis has made significant progress, driven by the success of deep learning techniques. Consequently, an increasing number of researchers are working to model the connection between imaging features and survival events.

Radiology involves the use of medical imaging technologies such as X-rays, CT (Computerized Tomography) scans, MRI (Magnetic Resonance Imaging) scans, and ultrasound to visualize internal structures and detect abnormalities. Radiological images can provide valuable macroscopic information such as lesion location, morphological texture, and tumor metastasis, which can help predict the prognosis for the patient~\cite{francone2020chest, swift2017magnetic, platz2017dynamic}. However, due to its lower sensitivity, radiology is not widely considered as the gold standard in cancer diagnosis. To improve diagnosis accuracy, the pathological examination will be conducted to sample lesion tissues and acquire pathological images, also known as whole slide images (WSIs). Pathological images can provide information about microscopic changes in tumor cells and their microenvironment. Generally, multi-instance learning (MIL) is the most commonly used paradigm in pathology-based survival analysis~\cite{zhu2017wsisa, wulczyn2020deep, yao2020whole, shao2021weakly, chen2022scaling}, which can identify and highlight important regions within the pathological image that contribute to the survival event, revealing insights into the underlying pathological phenotypes of disease. Recently, with the rapid advances in high-throughput sequencing technologies, more and more accessible large-scale \textit{genomics} datasets provide an unprecedented opportunity to deeply understand the survival events from the molecular perspective~\cite{lee2019review,dessie2021novel, dey2022efficient}.

Although survival analysis using single-modality has achieved promising results, combining multi-modal data from different perspectives can provide complementary information for each other. Intuitively, it can increase the sensitivity of survival analysis by detecting subtle changes that may not be visible within the single-modality. The most straightforward method is to integrate all features learned from multi-modal data together~\cite{cheerla2019deep, chen2020pathomic, braman2021deep}. Obviously, these methods neglect the potential correlations and interactions between multi-modalities, which is crucial for information sharing and feature fusion in medical image analysis. Thereby, the attention mechanism has been introduced to capture the shared context in multi-modalities. For example, some researchers utilized clinical reports~\cite{li2021multi, takagi2023transformer} or genomic profiles~\cite{wang2021gpdbn, chen2021multimodal} as guidance for models to focus on the relevant parts of pathological images. Under the supervision of advanced knowledge, models can identify useful phenotypes and discover possible biomarkers associated with specific gene expression or clinical outcomes.

The above-mentioned cross-modal interaction methods are plausible when the reference modality is superior to other retention modalities. However, in some cases, the performance of pathology-based survival analysis is better than genomics-based or report-based methods. In such cases, if we still leverage the worse modality as the reference to supervise the feature learning of better retention modalities, the more discriminative information in retention modalities will be contaminated by mediocre information in the reference modality. Moreover, the original purpose of multi-modal medical image analysis is to integrate the complementary information contained in multi-modal data and make more accurate predictions. These attention-based cross-modal interaction methods will discard pathological information irrelevant to gene expression or clinical reports.

In light of these observations, we propose a novel Cross-Modal Translation and Alignment (CMTA) framework to explore the intrinsic cross-modal correlations and transfer potential complementary information. Concretely, we construct two parallel encoder-decoder structures for multi-modal data to extract intra-modal representation within single-modality and generates cross-modal representation from cross-modal information. To explore the potential cross-modal correlations, we leverage a cross-modal attention module as the information bridge between different modalities to perform cross-modal interaction and transfer complementary information. The cross-modal representation is utilized to enhance and recalibrate intra-modal representation. Finally, all intra-modal representations are integrated to yield the final survival prediction. The main contributions of this paper can be summarized as follows:

\begin{itemize}
   \item We propose a novel Cross-Modal Translation and Alignment (CMTA) framework for survival analysis using pathological images and genomic profiles, where two parallel encoder-decoder structures are constructed for multi-modal data to integrate intra-modal information and generate cross-modal representation.
   \item We introduce the attention mechanism to design a cross-modal attention module, which is embedded into the encoder-decoder structure to explore the intrinsic cross-modal correlations, perform the potential interactions and transfer cross-modal complementary information between different modalities.
   \item We conduct extensive experiments on five public TCGA datasets to evaluate the effectiveness of our proposed model. The experimental results show that our model consistently achieves superior performance compared to the state-of-the-art methods.
\end{itemize}

\begin{figure*}
   \begin{center}
      \includegraphics[width=1.0\linewidth]{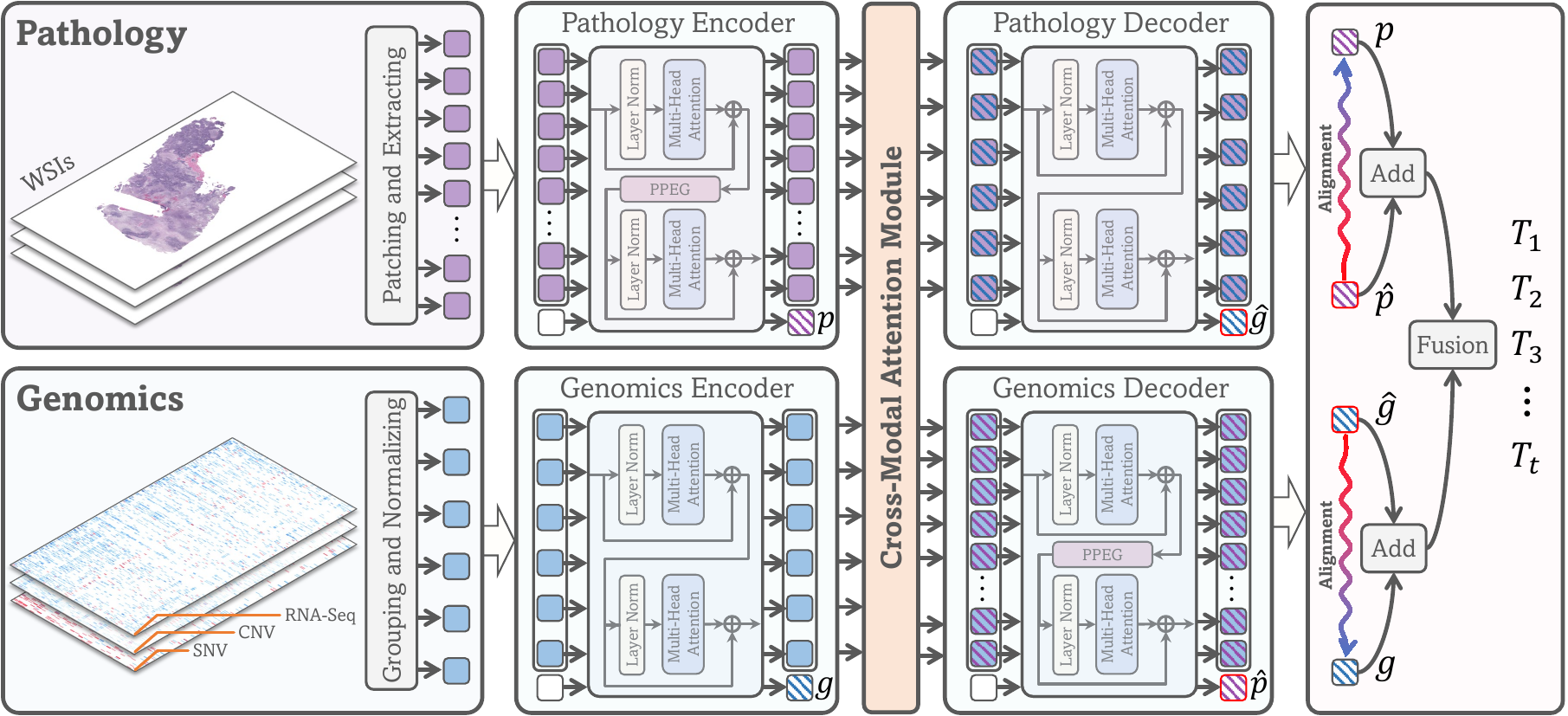}
   \end{center}
   \caption{Illustration of Cross-Modal Translation and Alignment (CMTA) framework. There are two parallel encoder-decoder structures, in which encoders are used to extract intra-modal representations ($p$ and $g$) and decoders are used to generate cross-modal representations ($\hat{p}$ and $\hat{g}$). There is a cross-modal attention module between encoder and decoder to explore intrinsic cross-modal correlations and transfer potential cross-modal information. Finally, intra-modal and cross-modal representations are fused to make survival predictions.}
   \label{framework}
\end{figure*}

\section{Related Work}
\subsection{Survival Analysis using Single-modality}
Survival prediction can provide valuable information for doctors to assess the clinical outcomes for disease progression and treatment efficiency. Traditional survival analysis relies on single-modal clinical data, such as short-term clinical indicators~\cite{hagar2014survival, kalafi2019machine, lai2020overall, yu2021dynamic}, long-term follow-up reports~\cite{ahearne2016short, capra2017assessing, vale2021long}, various radiological images~\cite{francone2020chest, swift2017magnetic, platz2017dynamic, van2017survival} and gigapixel pathological images~\cite{zhu2017wsisa, wulczyn2020deep, yao2020whole, chen2022scaling, shao2021weakly}. Typically, radiology-based methods would utilize feature learning techniques to extract quantitative features from radiological images and correlate them with survival outcomes. The pathology-based methods would leverage the MIL paradigm to can identify and highlight important instances (\textit{i.e.}, patches or regions) within an image that contribute to the survival event. Pathology can significantly improve the performance of survival prediction in comparison with radiology. However, the performance still cannot satisfy the requirements of clinical applications. With the rapid advances in high-throughput sequencing technologies, genomic profiles have shown high relevance as a measure for disease modeling and prognosis~\cite{christinat2015integrated, cheerla2017microrna, liu2017mirnas, yousefi2017predicting, qiu2020meta}. That opens a novel pathway for more accurate survival prediction.

\subsection{Survival Analysis using Multi-modality}
Multi-modal data can provide various insights into patient's condition at the macroscopic, microscopic, and molecular levels. Each modality has its strengths and weaknesses. By integrating information from different modalities, we can obtain more comprehensive understanding for the patient's condition, leading to more accurate diagnosis, treatment planning, and better prognosis prediction. For example, Cheerla \textit{et al.}~\cite{cheerla2019deep} developed an unsupervised method to encode multi-modal patient data into a common feature representation, and then used these feature representations to predict single-cancer and pan-cancer prognosis. Chen \textit{et al.}~\cite{chen2020pathomic} proposed a Pathomic Fusion framework to integrate histology images and genomic features for building objective image-omic assays for cancer diagnosis and prognosis. Braman \textit{et al.}~\cite{braman2021deep} presented a Deep Orthogonal Fusion model to encourage each modality to provide independent prognostic information. Chen \textit{et al.}~\cite{chen2021multimodal} proposed a Multimodal Co-Attention Transformer framework that identifies informative instances from pathological images using genomic features as queries. Although these methods have achieved impressive results in survival analysis using multi-modality, the straightforward fusion of multi-modal features would overlook potential cross-modal correlations, and using genomic profiles as guidance would discard pathological information irrelevant to gene expression. These methods violate the original purpose of integrating the complementary information contained in multi-modal data.

\section{Method}
In this section, we present the overall description of our proposed Cross-Modal Translation and Alignment (CMTA) framework for survival analysis, as illustrated in Figure \ref{framework}. First, we introduce the problem formulation for survival analysis incorporating pathological images with genomic profiles. Then, we detail the data processing and feature extraction for each modality. After that, we elaborate on the key components of our proposed framework one by one.

\subsection{Problem Formulation}
Let $\mathbb{X}=\{X_1, X_2, \cdots, X_N\}$ represent the clinical data of $N$ patients. Each patient data can be represented by a 4-tuples $X_i=(P_i, G_i, c_i, t_i)$, where $P_i$ is the set of whole slide images, $G_i$ is the set of genomic profiles,  $c_i\in \{0, 1\}$ is the right uncensorship status and $t_i\in \mathbb{R}^+$ is overall survival time (in months). In survival analysis, let $T$ be a continuous random variable for overall survival time. Our goal is to develop a survival prediction model $\mathcal{F}$ that integrates pathological images $P$ and genomic profiles $G$ to estimate the hazard function $f_{hazard}(T=t\vert T\geq t, X)$,
\begin{equation}
   f_{hazard}(T=t)=\lim\limits_{\partial t \to 0} \frac{P(t\leq T \leq t+ \partial t\vert T\geq t)}{\partial t}.
\end{equation}

The hazard function represents the instantaneous rate of occurrence of the event of interest at time $t$. In practical applications, we tend to measure the probability of patient surviving longer than a discrete time point $t$, rather than estimating survival time directly. The survival function can be obtained via the cumulative hazard function,
\begin{equation}
   f_{sur}(T\leq t, X)=\prod_{u=1}^t(1-f_{hazard}(T=u)).
\end{equation}

The hazard function can be estimated using various statistical models. The most common method for estimating hazard function is the Cox Proportional Hazards model~\cite{wang2019machine, wang2019extreme, zhu2016deep}, in which hazard function $f_{hazard}$ can be parameterized as an exponential linear function,
\begin{equation}
   h(t|X) = h_0(t)e^{\theta X},
\end{equation}
where $h_0(t)$ represents the baseline hazard function, $\theta$ represents the vector of coefficients for the covariates. Generally, $\theta$ is the learnable parameters of the last hidden layer in the neural network.

\subsection{Data processing and Feature Extraction}
\textbf{Pathological Images. }Following the previous works~\cite{chen2021multimodal, lu2021ai, shao2021transmil}, we adopt CLAM~\cite{lu2021data} to crop each WSI into a series of non-overlapping $512\times 512$ patches at $40\times$ magnification level. Then, ResNet-50 (pretrained on ImageNet) is used to extract $1024$-dim feature. Then, each patch feature is fed into a fully connected layer to obtain $d$-dimension embedding. For ease of notation, we drop $i$ in referring to the $i$-th patient. That means, the pathological images of each patient can be represented as $P=\{p_1, p_2,\cdots,p_M\}\in \mathbb{R}^{M\times d}$, where $M$ is the number of patches.

\textbf{Genomic Profiles. }Genomic profiles are the individual's most sensitive and identifiable information, including RNA sequencing (RNA-seq), Copy Number Variation (CNV), Simple Nucleotide Variation (SNV), DNA methylation, etc. Due to the higher signal-to-noise ratio, some of them have to be dropped in bioinformatics analysis. The genomic profiles used in this paper cover RNA-seq, CNV, and SNV. Following previous works~\cite{liberzon2015molecular, chen2021multimodal}, the genomic profiles can be grouped into the following genomic sequences: 1)Tumor Suppression, 2) Oncogenesis, 3) Protein Kinases, 4) Cellular Differentiation, 5) Transcription, and 6) Cytokines and Growth. Similar to pathomics, these grouped genomic sequences are fed into a fully connected layer to obtain $d$-dimension embeddings. That means, the genomics of each patient can be represented as $G=\{g_1, g_2,\cdots, g_K\}\in \mathbb{R}^{K\times d}$, where $K$ is the number of groups.

\subsection{Pathology Encoder and Genomics Encoder}
Recently, the self-attention mechanism has been proven to be one of the most powerful tools for integrating information and extracting features from the set-based data structure. Therefore, we introduce the self-attention mechanism to construct an encoder for each modality to integrate intra-modal information and obtain intra-modal representation. Note that the size of sets in this paper is extremely large, especially patch sets $P$. Traditional global self-attention will bring heavy computation burden. To tackle this issue, we utilize the Nystrom attention~\cite{xiong2021nystromformer} to approximate the global self-attention.

\textbf{Pathology Encoder. }For the given patch sets $P=\{p_1, p_2,\cdots,p_M\}$, we define a learnable class token $p^{(0)}$ to gather information from all patch features. The initial input of the pathology encoder is represented as $P^{(0)}=\{p^{(0)},p_1^{(0)}, p_2^{(0)},\cdots,p_M^{(0)}\}\in \mathbb{R}^{(M+1)\times d}$. We apply two self-attention layers to perform information integration. Additionally, there is another PPEG (Pyramid Position Encoding Generator.) module~\cite{shao2021transmil} to explore the correlations among different patches. The computation of the pathology encoder can be formulated as follows,
\begin{equation}
   P^{(1)}={\rm MSA}({\rm LN}(P^{(0)}))+P^{(0)},
\end{equation}
\begin{equation}
   P^{(2)}={\rm PPEG}(P^{(1)}),
\end{equation}
\begin{equation}
   P^{(3)}={\rm MSA}({\rm LN}(P^{(2)}))+P^{(2)},
\end{equation}
where MSA denotes Multi-head Self-attention and LN denotes Layer Norm. The output of this encoder is $P^{(3)}=\{p^{(3)},p_1^{(3)}, p_2^{(3)},\cdots,p_M^{(3)}\}\in \mathbb{R}^{(M+1)\times d}$. Let class token $p^{(3)}$ be the intra-modal representation of pathology, marking it as $p$.

\textbf{Genomics Encoder. }For the given genomics $G=\{g_1, g_2,\cdots,g_K\}$, we also define a learnable class token $g^{(0)}$ to gather information from all the gene sequences. The initial input of the genomics encoder is represented as $G^{(0)}=\{g^{(0)},g_1^{(0)}, g_2^{(0)},\cdots,g_K^{(0)}\}\in \mathbb{R}^{(K+1)\times d}$. The structure of genomics encoder is similar to pathology encoder with exception to PPEG module. The computation of the genomics encoder can be formulated as follows,
\begin{equation}
   G^{(1)}={\rm MSA}({\rm LN}(G^{(0)}))+G^{(0)},
\end{equation}
\begin{equation}
   G^{(2)}={\rm MSA}({\rm LN}(G^{(1)}))+G^{(1)}.
\end{equation}

Let $g^{(2)}$ in output $G^{(2)}=\{g^{(2)},g_1^{(2)}, g_2^{(2)},\cdots,g_K^{(2)}\}\in \mathbb{R}^{(K+1)\times d}$ be the intra-modal representation of genomics, marking it as $g$.

\subsection{Cross-Modal Attention Module}
\begin{figure}[t]
   \begin{center}
      \includegraphics[width=1.0\linewidth]{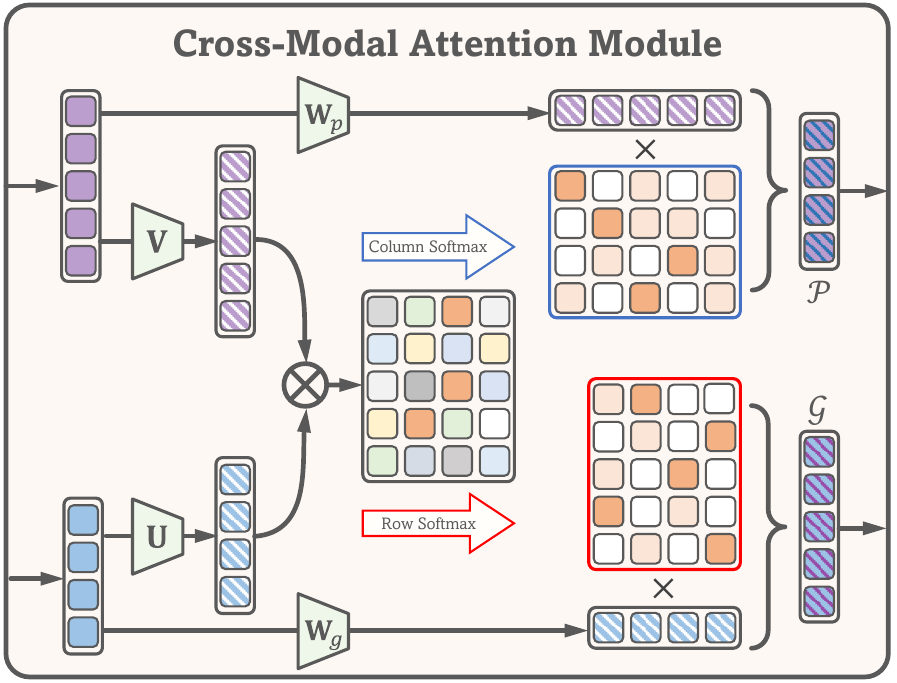}
   \end{center}
   \caption{The structure of cross-modal attention module. This module aims to explore potential correlations and interactions between multi-modal data.}
   \label{cross-modal}
\end{figure}

In this part, we denote the instance tokens of pathology encoder as $\slashed{P}=\{p_1^{(3)}, p_2^{(3)},\cdots,p_M^{(3)}\}\in \mathbb{R}^{M\times d}$, and the instance tokens of genomics encoder as $\slashed{G}=\{g_1^{(2)}, g_2^{(2)},\cdots,g_K^{(2)}\}\in \mathbb{R}^{K\times d}$. The cross-modal attention module is designed to explore the potential cross-modal correlations and interactions, as illustrated in Figure \ref{cross-modal}. This module takes $\slashed{P}$ and $\slashed{G}$ as input to calculate two attention maps $\mathcal{H}_p$ and $\mathcal{H}_g$,
\begin{equation}
   \begin{aligned}
      \mathcal{H}_p & ={\rm softmax}\left (\frac{(\slashed{G}{\rm {\textbf U}})\times (\slashed{P}{\rm {\textbf V}})^T}{\sqrt{d}}\right )                    \\
                    & ={\rm softmax}\left (\frac{\slashed{G}{\rm {\textbf U}}{\rm {\textbf V}}^T \slashed{P}^T}{\sqrt{d}}\right )\in \mathbb{R}^{K\times M},
   \end{aligned}
\end{equation}
\begin{equation}
   \begin{aligned}
      \mathcal{H}_g & ={\rm softmax}\left (\frac{(\slashed{P}{\rm {\textbf V}})\times (\slashed{G}{\rm {\textbf U}})^T}{\sqrt{d}}\right )                    \\
                    & ={\rm softmax}\left (\frac{\slashed{P}{\rm {\textbf V}}{\rm {\textbf U}}^T \slashed{G}^T}{\sqrt{d}}\right )\in \mathbb{R}^{M\times K},
   \end{aligned}
\end{equation}
where ${\rm {\textbf U}}\in \mathbb{R}^{d\times d}$ and ${\rm {\textbf V}}\in \mathbb{R}^{d\times d}$ are the learnable parameters. Essentially, the attention map $\mathcal{H}_p$ presents association status from genomics tokens to pathology tokens while the attention map $\mathcal{H}_g$ presents association status from pathology tokens to genomics tokens. With the help of attention maps, we can extract genomics-related information $\mathcal{P}$ in pathology tokens and pathology-related information $\mathcal{G}$ in genomics, respectively.
\begin{equation}
   \mathcal{P}=\mathcal{H}_p\times (\slashed{P}{\rm {\textbf W}}_p)=\mathcal{H}_p \slashed{P}{\rm {\textbf W}}_p \in \mathbb{R}^{K\times d},
\end{equation}
\begin{equation}
   \mathcal{G}=\mathcal{H}_g\times (\slashed{G}{\rm {\textbf W}}_g)=\mathcal{H}_g \slashed{G}{\rm {\textbf W}}_g \in \mathbb{R}^{M\times d},
\end{equation}
where ${\rm {\textbf W}}_p\in \mathbb{R}^{d\times d}$ and ${\rm {\textbf W}}_g\in \mathbb{R}^{d\times d}$ are the learnable parameters. In such manner, we can explore the potential cross-modal relationships and transfer complementary information between multi-modal data.

\subsection{Genomics Decoder and Pathology Decoder}
Due to the data heterogeneity gap between pathological images and genomic profiles, it is unreasonable to directly superpose $\mathcal{P}$ and $\mathcal{G}$ onto intra-modal representations $p$ and $g$. Therefore, we construct two decoders, i.e., pathology decoder and genomics decoder, to translate related information into specific cross-modal representations. For simplicity, pathology decoder has the same structure as genomics encoder while genomics decoder has the same structure as pathology encoder.

\textbf{Pathology Decoder. } For the genomics-related information $\mathcal{P}=\{{\rho}_1, {\rho}_2,\cdots, {\rho}_K\}$ in pathology, we define a learnable class token ${\rho}^{(0)}$, and then apply two self-attention layers to perform information translation.
\begin{equation}
   \mathcal{P}^{(1)}={\rm MSA}({\rm LN}(\mathcal{P}^{(0)}))+\mathcal{P}^{(0)}
\end{equation}
\begin{equation}
   \mathcal{P}^{(2)}={\rm MSA}({\rm LN}(\mathcal{P}^{(1)}))+\mathcal{P}^{(1)}
\end{equation}

Let ${\rho}^{(2)}$ in output $\mathcal{P}^{(2)}=\{{\rho}^{(2)},{\rho}_1^{(2)}, {\rho}_2^{(2)},\cdots,{\rho}_K^{(2)}\}\in \mathbb{R}^{(K+1)\times d}$ be the cross-modal representation learned from pathology, marking it as $\hat{g} $.

\textbf{Genomics Decoder. } For the pathology-related information $\mathcal{G}=\{{\xi}_1, {\xi}_2,\cdots, {\xi}_M\}$ in genomics, we also define a learnable class token ${\xi}^{(0)}$, and apply two self-attention layers with PPEG module to perform information translation.

\begin{equation}
   \mathcal{G}^{(1)}={\rm MSA}({\rm LN}(\mathcal{G}^{(0)}))+\mathcal{G}^{(0)},
\end{equation}
\begin{equation}
   \mathcal{G}^{(2)}={\rm PPEG}(\mathcal{G}^{(1)}),
\end{equation}
\begin{equation}
   \mathcal{G}^{(3)}={\rm MSA}({\rm LN}(\mathcal{G}^{(2)}))+\mathcal{G}^{(2)}.
\end{equation}

Let ${\xi}^{(3)}$ in output $\mathcal{G}^{(3)}=\{{\xi}^{(3)},{\xi}_1^{(3)}, {\xi}_2^{(3)},\cdots,{\xi}_M^{(3)}\}\in \mathbb{R}^{(M+1)\times d}$ be the cross-modal representation learned from genomics, marking it as $\hat{p}$.

\begin{table*}
   \small
   \begin{center}
      \begin{tabular}{lccccccc}
         \toprule
         {\multirow{2}{*}{Methods}}        & \multicolumn{2}{c}{Modality} & \multicolumn{5}{c}{Datasets}                                                                                                                                                                                      \\ \cmidrule(lr){2-3} \cmidrule(lr){4-8}
                                           & P.                           & G.                           & BLCA                              & BRCA                              & GBMLGG                            & LUAD                              & UCEC                               \\
         \midrule
         SNN~\cite{klambauer2017self}      &                              & $\checkmark$                 & 0.6339 $\pm$ 0.0509               & 0.6327 $\pm$ 0.0739               & 0.8370 $\pm$ 0.0276               & 0.6171 $\pm$ 0.0411               & 0.6900 $\pm$ 0.0389                \\
         SNNTrans~\cite{klambauer2017self} &                              & $\checkmark$                 & 0.6456 $\pm$ 0.0428               & 0.6478 $\pm$ 0.0580               & 0.8284 $\pm$ 0.0158               & 0.6335 $\pm$ 0.0493               & 0.6324 $\pm$ 0.0324                \\
         \midrule
         MaxMIL                            & $\checkmark$                 &                              & 0.5509 $\pm$ 0.0315               & 0.5966 $\pm$ 0.0547               & 0.7136 $\pm$ 0.0574               & 0.5958 $\pm$ 0.0600               & 0.5626 $\pm$ 0.0547                \\
         MeanMIL                           & $\checkmark$                 &                              & 0.5847 $\pm$ 0.0324               & 0.6110 $\pm$ 0.0286               & 0.7896 $\pm$ 0.0367               & 0.5763 $\pm$ 0.0536               & 0.6653 $\pm$ 0.0457                \\
         AttMIL~\cite{ilse2018attention}   & $\checkmark$                 &                              & 0.5673 $\pm$ 0.0498               & 0.5899 $\pm$ 0.0472               & 0.7974 $\pm$ 0.0336               & 0.5753 $\pm$ 0.0744               & 0.6507 $\pm$ 0.0330                \\
         CLAM-SB~\cite{lu2021data}         & $\checkmark$                 &                              & 0.5487 $\pm$ 0.0286               & 0.6091 $\pm$ 0.0329               & 0.7969 $\pm$ 0.0346               & 0.5962 $\pm$ 0.0558               & 0.6780 $\pm$ 0.0342                \\
         CLAM-MB~\cite{lu2021data}         & $\checkmark$                 &                              & 0.5620 $\pm$ 0.0313               & 0.6203 $\pm$ 0.0520               & 0.7986 $\pm$ 0.0320               & 0.5918 $\pm$ 0.0591               & 0.6821 $\pm$ 0.0646                \\
         TransMIL~\cite{shao2021transmil}  & $\checkmark$                 &                              & 0.5466 $\pm$ 0.0334               & 0.6430 $\pm$ 0.0368               & 0.7916 $\pm$ 0.0272               & 0.5788 $\pm$ 0.0303               & 0.6799 $\pm$ 0.0304                \\
         \midrule
         MCAT~\cite{chen2021multimodal}    & $\checkmark$                 & $\checkmark$                 & {\color{blue}0.6727 $\pm$ 0.0320} & 0.6590 $\pm$ 0.0418               & 0.8350 $\pm$ 0.0233               & 0.6597 $\pm$ 0.0279               & 0.6336 $\pm$ 0.0506                \\
         M3IF~\cite{li2021multi}           & $\checkmark$                 & $\checkmark$                 & 0.6361 $\pm$ 0.0197               & 0.6197 $\pm$ 0.0707               & 0.8238 $\pm$ 0.0170               & 0.6299 $\pm$ 0.0312               & 0.6672 $\pm$ 0.0293                \\
         GPDBN~\cite{wang2021gpdbn}        & $\checkmark$                 & $\checkmark$                 & 0.6354 $\pm$ 0.0252               & 0.6549 $\pm$ 0.0332               & {\color{blue}0.8510 $\pm$ 0.0243} & 0.6400 $\pm$ 0.0478               & 0.6839 $\pm$ 0.0529                \\
         Porpoise~\cite{chen2022pan}       & $\checkmark$                 & $\checkmark$                 & 0.6461 $\pm$ 0.0338               & 0.6207 $\pm$ 0.0544               & 0.8479 $\pm$ 0.0128               & 0.6403 $\pm$ 0.0412               & {\color{blue} 0.6918 $\pm$ 0.0488} \\
         HFBSurv~\cite{li2022hfbsurv}      & $\checkmark$                 & $\checkmark$                 & 0.6398 $\pm$ 0.0277               & 0.6473 $\pm$ 0.0346               & 0.8383 $\pm$ 0.0128               & 0.6501 $\pm$ 0.0495               & 0.6421 $\pm$ 0.0445                \\
         \midrule
         DualTrans                         & $\checkmark$                 & $\checkmark$                 & 0.6607 $\pm$ 0.0319               & {\color{blue}0.6637 $\pm$ 0.0621} & 0.8393 $\pm$ 0.0174               & {\color{blue}0.6706 $\pm$ 0.0343} & 0.6724 $\pm$ 0.0192                \\
         \rowcolor{dino} CMTA (Ours)       & $\checkmark$                 & $\checkmark$                 & {\color{red}0.6910 $\pm$ 0.0426}  & {\color{red}0.6679 $\pm$ 0.0434}  & {\color{red}0.8531 $\pm$ 0.0116}  & {\color{red}0.6864 $\pm$ 0.0359}  & {\color{red}0.6975 $\pm$ 0.0409}   \\
         \bottomrule
      \end{tabular}
   \end{center}
   \caption{The performance of different approaches on five public TCGA datasets. ``P.'' indicates whether to use pathological images and ``G.'' indicates whether to use genomic profiles. The best and second best results are highlighted in {\color{red} red} and {\color{blue} blue}, respectively.}
   \label{tab:main}
\end{table*}

\subsection{Feature Alignment and Fusion}
The cross-modal representations can provide complementary information that may not be visible within single-modality. Therefore, we utilize the cross-modal representations to enhance and recalibrate intra-modal representations. And then, all feature representations are integrated to yield the final survival prediction. Note that, gigapixel pathological images cause that the model cannot be optimized with mini-batch manner. The alternative optimization strategy is to consider discrete time intervals and model each interval using an independent output. The feature fusion and survival prediction can be formulated as,
\begin{equation}
   T_1,\cdots,T_t={\rm sigmoid}\left({\rm MLP}\left(\frac{p+\hat{p}}{2}\oplus \frac{g+\hat{g}}{2} \right)\right)
\end{equation}
where $\oplus$ denotes the concatenation operation. In practice, professional pathologists and biologists can estimate partial gene expression from pathological images or imagine possible pathological phenotypes from genomic profiles. Our decoders are designed for simulating this process to translate cross-modal information. To ensure the quality of information translation, we must impose alignment constraints on cross-modal representations. In this paper, we utilize $L_1$ norm to measure the distance between cross-modal representations and intra-modal representations,
\begin{equation}
   \mathcal{L}_{sim} = \frac{1}{d}\left(||p-\hat{p}||_1+||g-\hat{g}||_1\right).
\end{equation}

Note that, intra-modal representations $p$ and $g$ MUST be detached from the computational graph when we calculate $\mathcal{L}_{sim}$. That means this alignment optimization objective is unidirectional ($\hat{p}\rightarrow p$ and $\hat{g}\rightarrow g$). Otherwise, this model will converge to learn redundant shared information and fail to predict survival events.

We leverage NLL (negative log-likelihood) survival loss~\cite{chen2021multimodal} as the loss function of the survival prediction part. Unifying these two losses, we can obtain the total loss function of our CMTA framework,
\begin{equation}
   \mathcal{L}_{total} = \mathcal{L}_{sur} + \alpha \mathcal{L}_{sim},
\end{equation}
where $\alpha$ is a positive hyper-parameter for reconciling the contribution of alignment loss function.

\section{Experiments}
In this section, we conduct extensive experiments on five public TCGA datasets to evaluate the effectiveness of our model. We first introduce the datasets and evaluation metrics used in our study. Then, experimental results are compared with some state-of-the-art methods to demonstrate the superiority of our model. Finally, we conduct ablation studies to discuss the impacts of some key components.
\subsection{Datasets and Evaluation Metrics}
\textbf{Datasets. }The Cancer Genome Atlas (TCGA) \footnotemark[3] is a public database that contains genomic and clinical data from thousands of cancer patients, covering 33 types of common cancer. It has been used extensively in survival analysis to identify genetic alterations and molecular pathways associated with cancer survival. In this paper, we use prognosis data of five cancer datasets to evaluate our model, including Bladder Urothelial Carcinoma (BLCA) ($n=373$), Breast Invasive Carcinoma (BRCA) ($n=956$), Glioblastoma \& Lower Grade Glioma (GBMLGG) ($n=569$), Lung Adenocarcinoma (LUAD) ($n=453$) and Uterine Corpus Endometrial Carcinoma (UCEC) ($n=480$). For each dataset, we adopt 5-fold cross-validation splits to evaluate our model and other comparison methods.

\footnotetext[3]{\href{https://portal.gdc.cancer.gov}{https://portal.gdc.cancer.gov}}

\textbf{Evaluation Metrics. } The c-index, also known as the concordance index, is a metric used to evaluate the performance of survival analysis models. It measures the ability of a model to correctly order pairs of individuals in terms of their predicted survival times. The c-index can be formulated as follows,
\begin{equation}
   {\rm c\text{-}index}=\frac{1}{n(n-1)}\sum\limits_{i=1}^{n}\sum\limits_{j=1}^{n}I(T_i<T_j)(1-c_j)
\end{equation}
where $n$ is the number of cases, $T_i$ and $T_j$ are the survival times of $i$-th patient and $j$-th patient. $I(\cdot)$ is the indicator function, which takes the value 1 if its argument is true, and 0 otherwise. $c_j$ is the right censorship status.

\begin{table*}
   \small
   \begin{center}
      \begin{tabular}{lcccccc}
         \toprule
         {\multirow{2}{*}{Modules or Constraints}} & \multicolumn{5}{c}{Datasets}                                                                                                                                                 \\\cmidrule(lr){2-6}
                                                   & BLCA                             & BRCA                             & GBMLGG                           & LUAD                             & UCEC                             \\
         \midrule

         w/o Cross-Modal Attention                 & 0.6784 $\pm$ 0.0276              & 0.6397 $\pm$ 0.0612              & 0.8489 $\pm$ 0.0154              & 0.6371 $\pm$ 0.0245              & 0.6679 $\pm$ 0.0446              \\
         w/o Alignment Constraints                 & 0.6730 $\pm$ 0.0209              & 0.6304 $\pm$ 0.0367              & 0.8473 $\pm$ 0.0200              & 0.6764 $\pm$ 0.0211              & 0.6643 $\pm$ 0.0500              \\
         w/o Tensor Detaching                      & 0.6002 $\pm$ 0.0501              & 0.6416 $\pm$ 0.0486              & 0.8256 $\pm$ 0.0248              & 0.6399 $\pm$ 0.0315              & 0.6504 $\pm$ 0.0153              \\
         w/o PPEG module                           & 0.6629 $\pm$ 0.0162              & 0.6627 $\pm$ 0.0423              & {\color{red}0.8582 $\pm$ 0.0194} & 0.6815 $\pm$ 0.0337              & 0.6659 $\pm$ 0.0509              \\
         \rowcolor{dino} CMTA (All Components)     & {\color{red}0.6910 $\pm$ 0.0426} & {\color{red}0.6679 $\pm$ 0.0434} & 0.8531 $\pm$ 0.0116              & {\color{red}0.6864 $\pm$ 0.0359} & {\color{red}0.6975 $\pm$ 0.0409} \\

         \bottomrule
      \end{tabular}
   \end{center}
   \caption{The experimental results after removing three key components: 1) Removing the cross-modal attention module; 2) Not imposing the alignment constraints; and 3) Not detaching $p$ and $g$ when calculating $\mathcal{L}_{sim}$.}
   \label{tab:ablation}
\end{table*}



\subsection{Comparisons with State-of-the-Art}
To perform the more comprehensive comparison, we implemented and evaluated some latest survival prediction methods using the same 5-fold cross-validation splits. These methods cover the single-modal learning paradigm and multi-modal learning paradigm. Table \ref{tab:main} shows the experimental results of all methods on all five TCGA datasets. It is worth noting that some of these methods can be regarded as the baseline of our model.

\textbf{Baseline Models. }1) SNNTrans: This model is the variation of SNN (Self-Normalizing Network)~\cite{klambauer2017self}, where we apply the same self-attention structure with the genomics encoder to integrate genomic information. This model is the single-modal baseline using genomic profiles. 2) TransMIL~\cite{shao2021transmil}: This method is one of the state-of-the-art MIL frameworks, which has achieved superior results on some public WSI classification benchmarks. In our study, we modify its classifier to solve the survival prediction task. This model is the single-modal baseline using pathological images. 3) DualTrans: This model is derived from SNNTrans and TransMIL, which concatenates the intra-modal representations learned by SNNTrans and TransMIL to make survival predictions. This model is the multi-modal baseline using pathomics and genomics.




\textbf{Compared with Single-modal Models. }As we can see from Table \ref{tab:main}, our proposed method consistently achieves superior performance on all TCGA datasets. More concretely, our model obtains c-index of 69.10\% on BLCA, 66.79\% on BRCA, 85.31\% on GBMLGG, 68.64\% on LUAD, and 69.75\% on UCEC, improving over the previous best single-modal methods by 4.54\%, 2.01\%, 1.61\%, 5.29\% and 0.75\%, respectively. This comparison results also show the advantages of survival prediction using multi-modality.

\textbf{Compared with Multi-modal Models. }The MCAT is the previous state-of-the-art multi-modal method, which leverages genomics as guidance to integrate pathological information. Against MCAT, our model achieves the performance increases of 1.83\% on BLCA, 0.89\% on BRCA, 1.81\% on GBMLGG, 2.67\% on LUAD and 6.39\% on UCEC. Note that MCAT would discard the discard pathological information irrelevant to gene expression while our method fully exploits all information contained in multi-modal data. That means survival analysis using multi-modality should focus on integrating complementary information between different modalities, rather than exploiting the abundant shared information. Besides, our model also consistently outperforms other SOTA multi-modal learning method by a large margin, including M3IF~\cite{li2021multi}, GPDBN~\cite{wang2021gpdbn}, Porpoise~\cite{chen2022pan} and HFBSurv~\cite{li2022hfbsurv}.

\textbf{Compared with Baseline Models. }From observations, it is obvious that DualTrans concatenating the intra-modal representations learned by SNNTrans and TransMIL can significantly improve the performance. It also shows the advantage of multi-modal data for accurate survival prediction. In our model, we utilize the cross-modal representations to enhance and recalibrate intra-modal representations. Then, we adopt the same feature fusion strategy with DualTrans to yield final survival prediction. Compared with DualTrans, our model achieves the performance increases of 3.03\% on BLCA, 0.42\% on BRCA, 1.38\% on GBMLGG, 1.58\% on LUAD and 2.51\% on UCEC. The improvements demonstrate that it is effective to translate related information into specific cross-modal representations.

\begin{figure}[t]
   \begin{center}
      \includegraphics[width=0.95\linewidth]{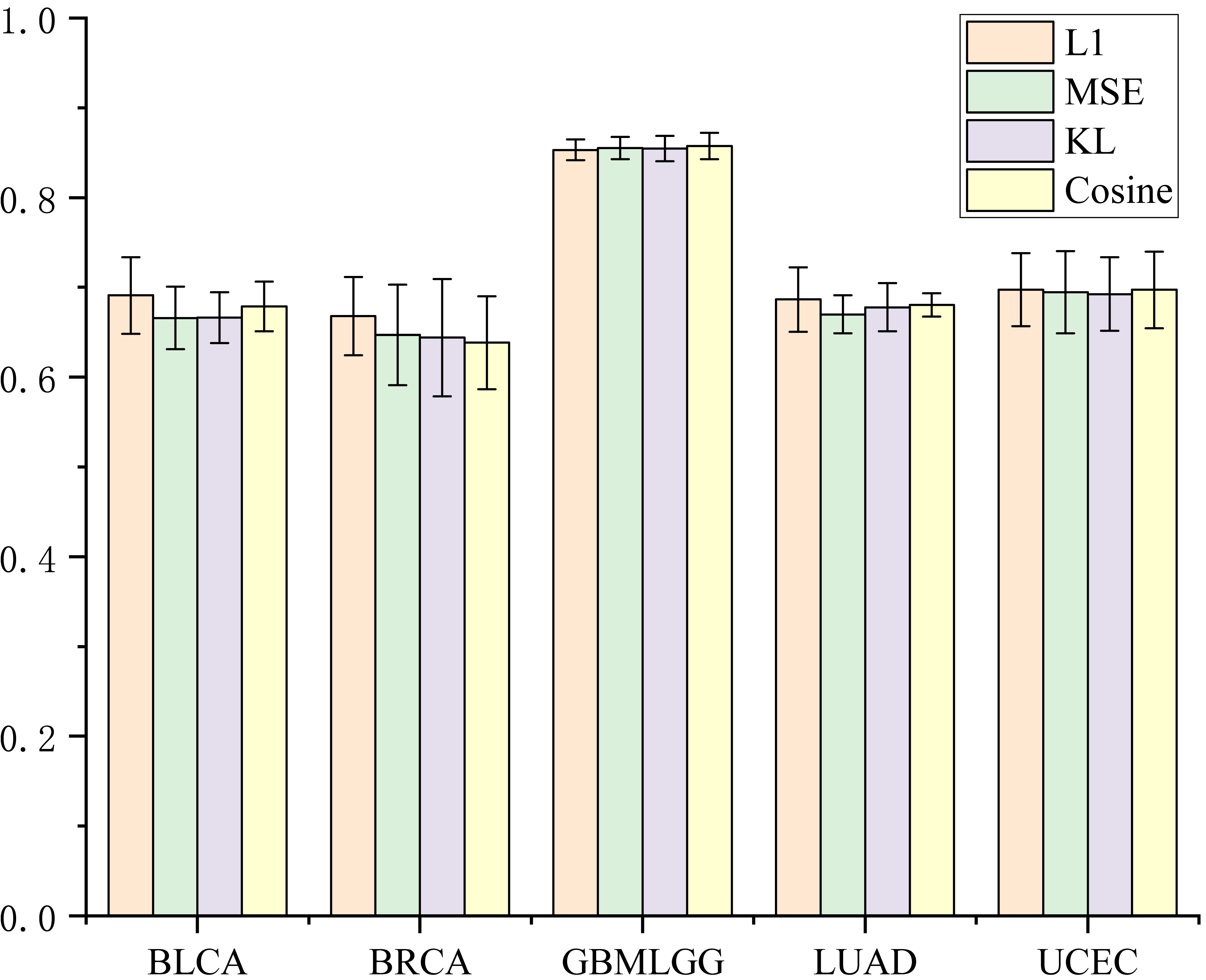}
   \end{center}
   \caption{Performance under different similarity metrics.}
   \label{metrics}
\end{figure}

\begin{figure*}[h]
   \centering
   \hspace{-3mm}
   \subfigure[BLCA]{
      \includegraphics[width=0.2\linewidth]{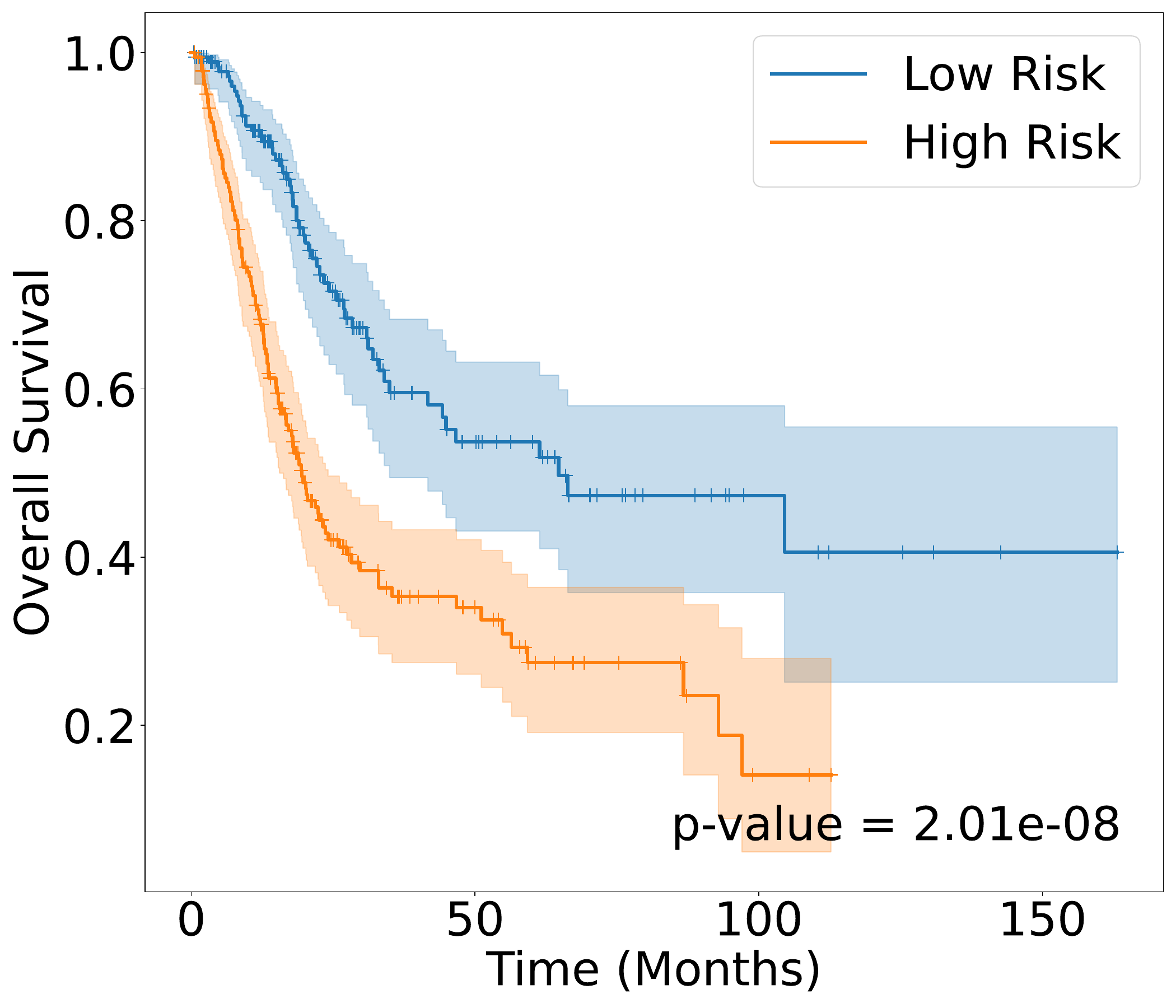}
      \label{BLCA}
   }\hspace{-3mm}
   \subfigure[BRCA]{
      \includegraphics[width=0.2\linewidth]{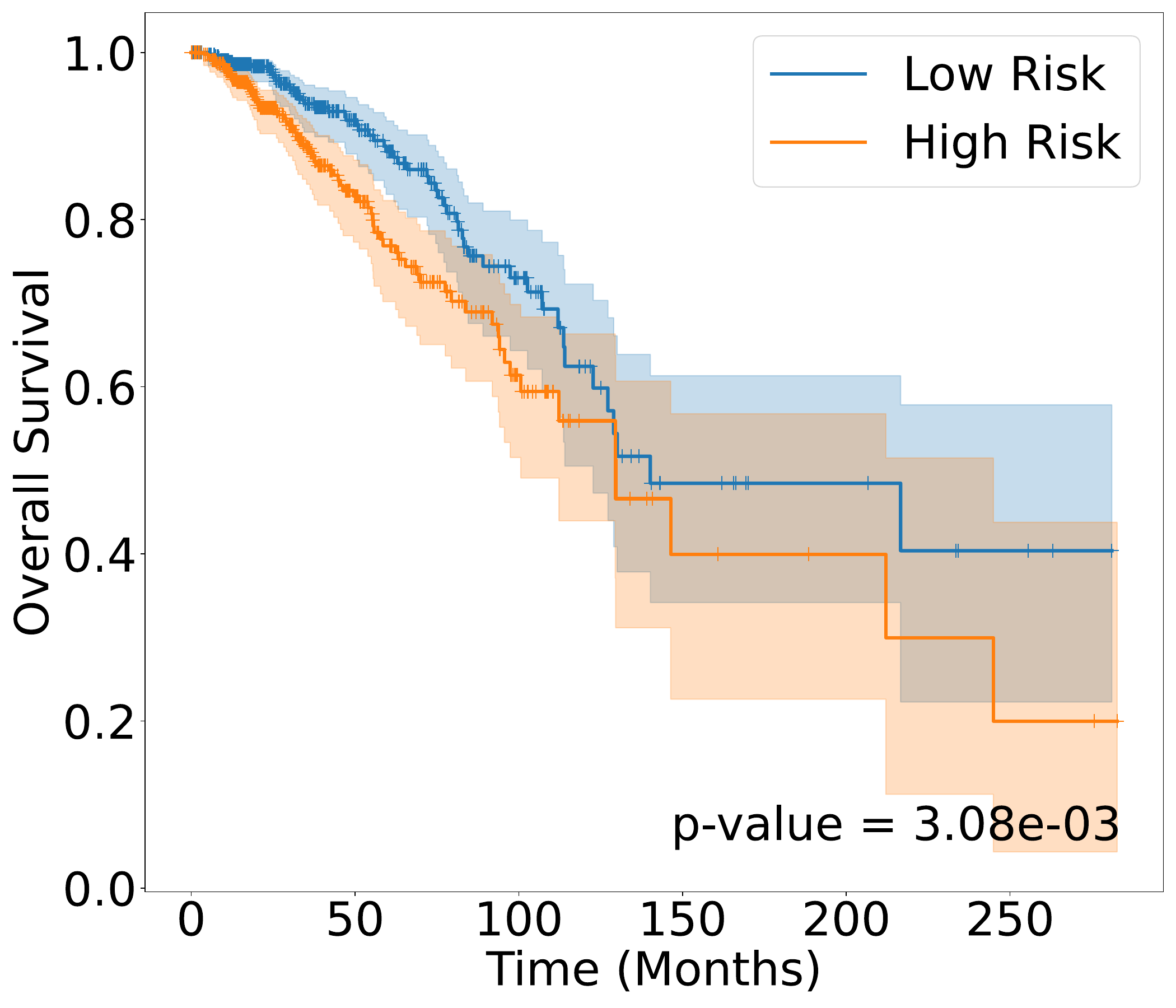}
      \label{BRCA}
   }\hspace{-3mm}
   \subfigure[GBMLGG]{
      \includegraphics[width=0.2\linewidth]{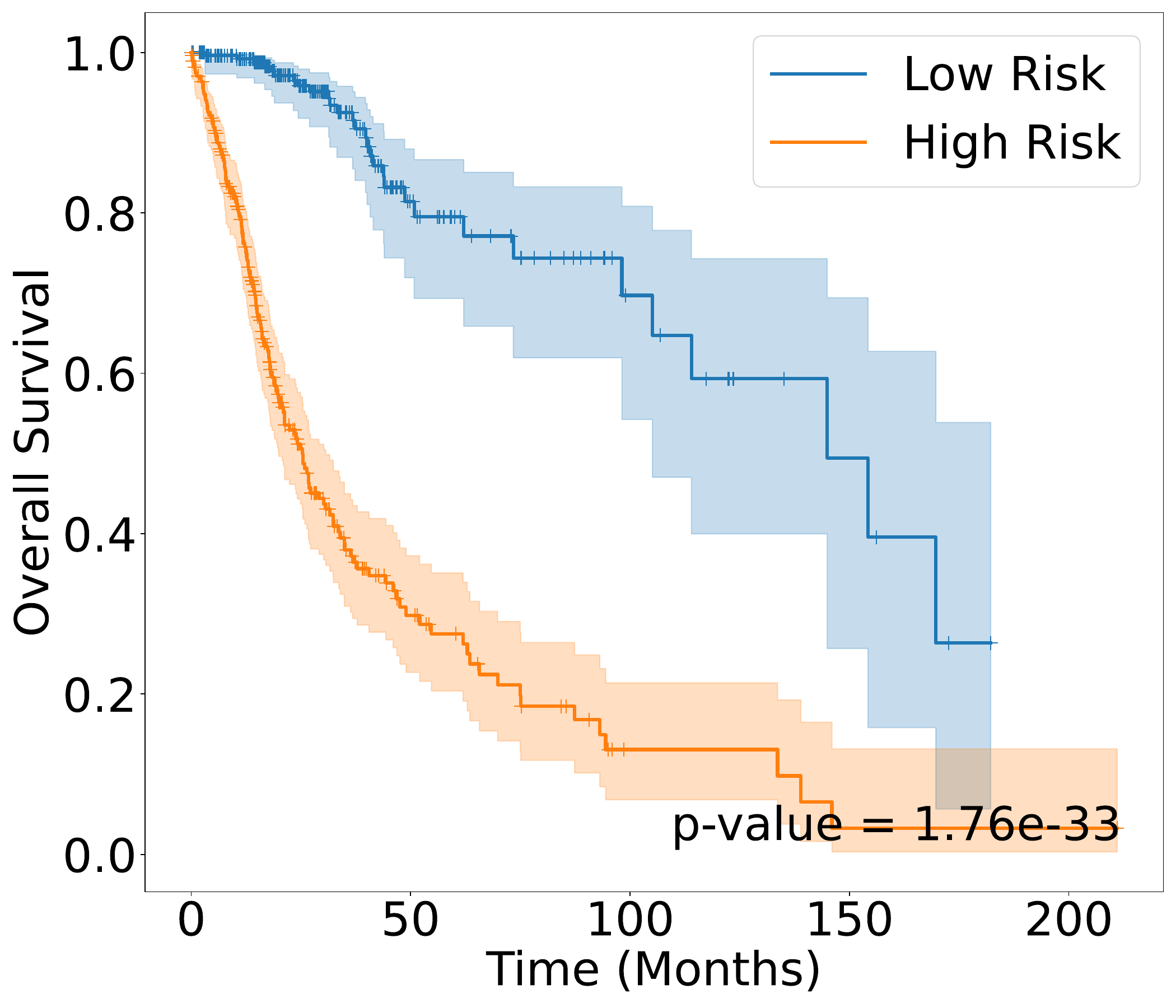}
      \label{GBMLGG}
   }\hspace{-3mm}
   \subfigure[LUAD]{
      \includegraphics[width=0.2\linewidth]{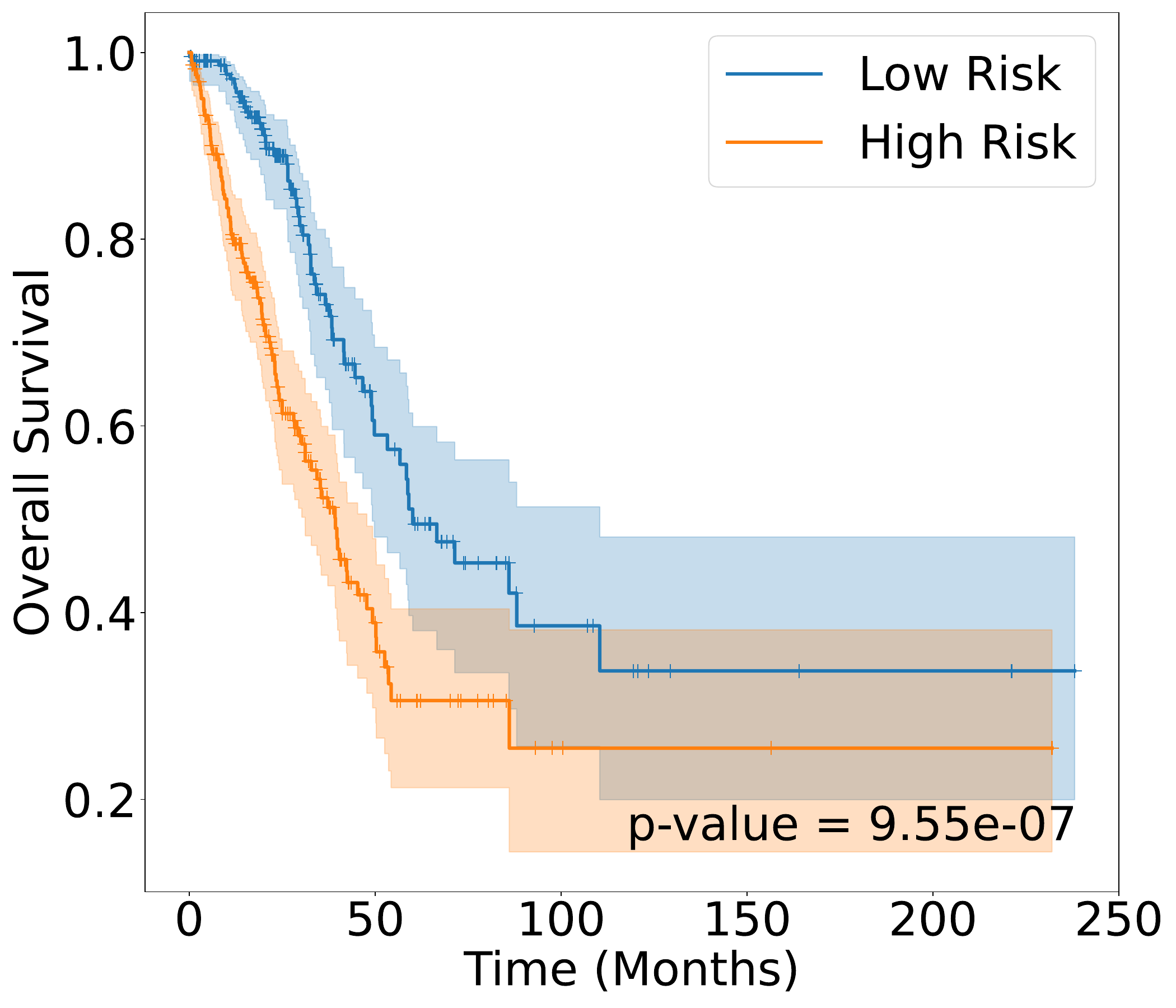}
      \label{LUAD}
   }\hspace{-3mm}
   \subfigure[UCEC]{
      \includegraphics[width=0.2\linewidth]{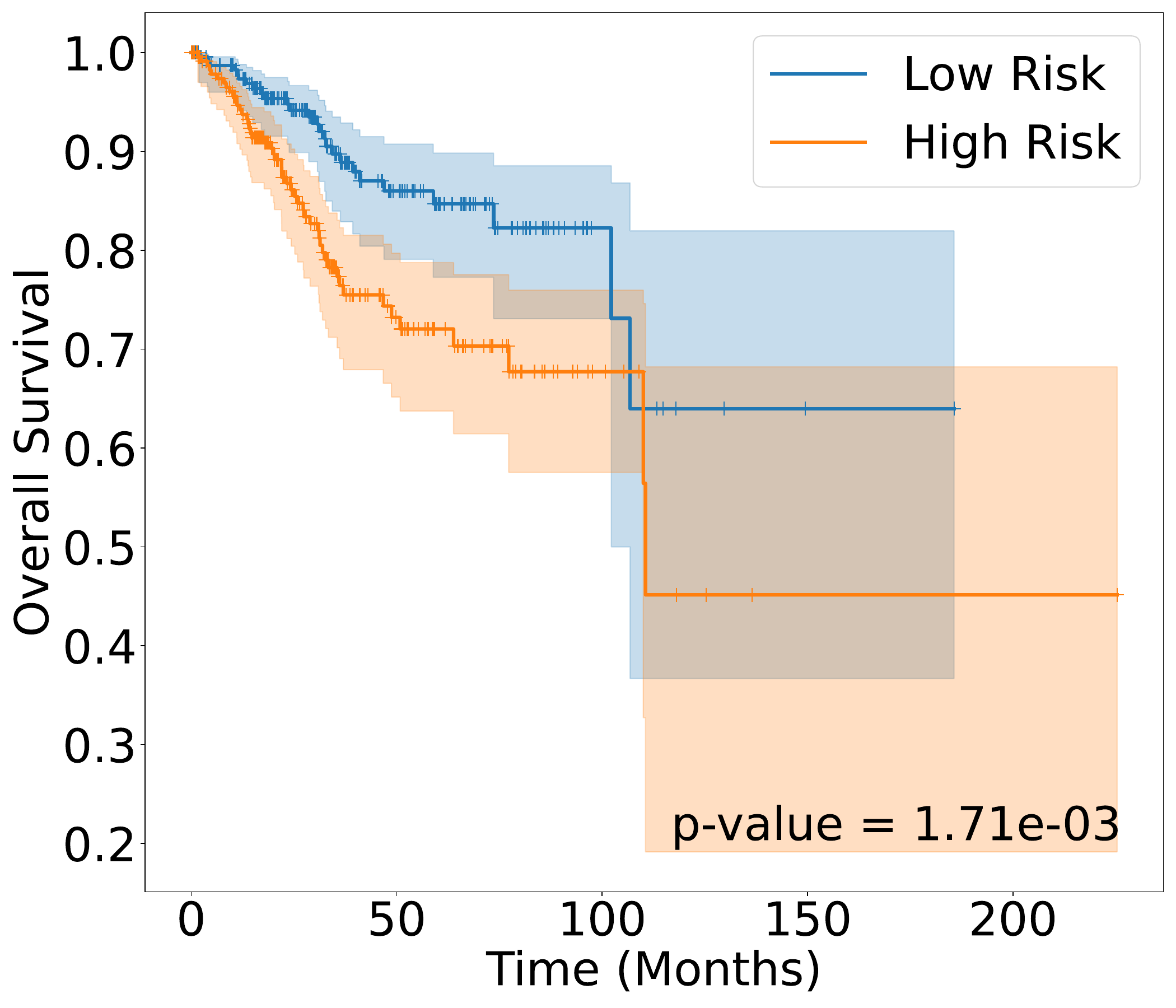}
      \label{UCEC}
   }\hspace{-3mm}
   \caption{According to predicted risk scores, all patients are stratified into low risk group (blue) and high risk group (red). Then, we utilize Kaplan-Meier analysis and Logrank test ($p$-value) to measure the statistical significance between low risk group and high risk group.}
   \label{KM}
\end{figure*}

\subsection{Ablation Studies}
In this part, we conduct some extra experiments to further discuss the impacts of different similarity metrics, components, and constraints.

\textbf{Impacts of Similarity Metrics. }By default, we leverage $L_1$ norm as the similarity metric to measure the distance between intra-modal representation and cross-modal representation. In this part, we conduct some experiments to evaluate the performances under other common similarity metrics, i.e., MSE (mean squared error) loss, KL (Kullback-Leibler) divergence, and Cosine similarity. The experimental results are shown in Figure~\ref{metrics}. As we can see from this figure, $L_1$ loss consistently outperforms other similarity metrics with the exception of GBMLGG dataset. Specifically, there are no obvious performance differences on GBMLGG dataset between different similarity metrics. That is because glioma is a very special cancer, which has been proven that five-year and ten-year survival rates are highly correlated with some specific genes~\cite{tykocki2018ten, jovvcevska2019genetic, basindwah2022ten}. Incorporating pathological images with genomic profiles may not change these specific gene expressions.

\textbf{Impacts of Modules and Constraints. } We remove some modules and constraints to investigate their impacts. The experimental results are summarized in Table~\ref{tab:ablation}. 1) The cross-modal attention module aims to explore potential correlations and interactions between multi-modal data. After removing this module, the performance drops 1.26\% on BLCA, 2.82\% on BRCA, 0.42\% on GBMLGG, 4.93\% on LUAD and 2.96\% on UCEC, which implies that highlighting related information is necessary when performing information translation. 2) To ensure the quality of information translation, we impose alignment constraints $\mathcal{L}_{sim}$ on cross-modal representations. If we remove this penalty term in model optimization, the performance will lose 1.80\% on BLCA, 3.75\% on BRCA, 0.58\% on GBMLGG, 1.00\% on LUAD, and 3.32\% on UCEC, respectively. That is, the unconstrained information translation process will severely impair the discrimination of intra-modal representations and deteriorate the performance of survival prediction. 3) To ensure the unidirectional alignment optimization objective, we detach intra-modal representations $p$ and $g$ from the computational graph when calculating loss function $\mathcal{L}_{sim}$. If we remove this detaching step, the performance will also significantly lose 9.08\% on BLCA, 2.63\% on BRCA, 2.75\% on GBMLGG, 4.65\% on LUAD and 4.71\% on UCEC, respectively. Without the detaching step, our model will converge to learn redundant shared information and fail to predict survival events. The degree of performance deterioration relies on the proportion of shared information in multi-modal data, and less shared information would result in more performance drop. 4) PPEG is one of key modules in TransMIL, used to explore the position correlations among different patches. Removing PPEG module will result in performance drops of 2.81\% on BLCA, 0.52\% on BRCA, 0.49\% on LUAD and 3.16\% on UCEC.

\subsection{Survival Analysis}
To further validate the effectiveness of CMTA for survival analysis, we stratify all patients into low risk group and high risk group according mid-value of the predicted risk scores from CMTA. After that, we utilize Kaplan-Meier analysis to visualize the survival events of all patients, analysis results are shown in Figure~\ref{KM}. Meanwhile, we also utilize Logrank test ($p$-value) to measure the statistical significance between low risk group (blue) and high risk group (red). A $p$-value of 0.05 or lower is generally considered statistically significant. As we can see from this figure, $p$-value of all datasets are significantly smaller than 0.05.



\section{Conclusion}
In this paper, we proposed a novel Cross-Modal Translation and Alignment (CMTA) framework for survival analysis using pathological images and genomic profiles, in which two parallel encoder-decoder structures are constructed for pathological features and genomic features to integrate intra-modal information and generate cross-modal representations, respectively. To explore the potential cross-modal correlations and interactions, we designed a cross-modal attention module as the information bridge between different modalities. Using cross-modal representations to enhance and recalibrate intra-modal representations can significantly improve the performance of survival prediction. Extensive experimental results on five public TCGA datasets demonstrated the effectiveness of our proposed model over state-of-the-art methods.

\section{Acknowledgement}
This work was supported by National Natural Science Foundation of China (No.~62202403), the Research Grants Council of the Hong Kong Special Administrative Region, China (No.~R6003-22) and Hong Kong Innovation and Technology Fund (No.~PRP/034/22FX).

   {\small
      \bibliographystyle{ieee_fullname}
      \bibliography{egbib}
   }

\end{document}